\documentclass[11pt,onecolumn]{article}

\usepackage{amsfonts}
\usepackage{amsmath}
\usepackage{amssymb}
\usepackage{cite}
\usepackage[amsmath,thmmarks]{ntheorem}
\usepackage[dvips]{graphicx}
\usepackage{subfigure}
\usepackage{color, soul}
\usepackage{epic, eepic}
\usepackage[top=1in, bottom=1in, left=1.25in, right=1.25in]{geometry}
\usepackage{appendix}

\usepackage{theorem}

\theoremheaderfont{\sc}\theorembodyfont{\upshape}
\theoremstyle{nonumberplain}
\theoremseparator{}
\theoremsymbol{\rule{1ex}{1ex}}

\usepackage{array}

\newcommand{\bm}[1]{\mbox{\boldmath{$#1$}}}

\begin{document}
\title{A Robust Zero-point Attraction LMS Algorithm on Near Sparse System Identification }

\author{Jian Jin,~
        Qing Qu,~
        and~Yuantao~Gu\thanks{This work was partially supported by National Natural Science Foundation of China (NSFC 60872087 and NSFC U0835003).
        The authors are with Department of Electronic Engineering, Tsinghua University, Beijing 100084, China. The corresponding author of this paper is Yuantao Gu (Email: gyt@tsinghua.edu.cn).}}

\date{Received Feb. 2012; accepted Feb. 2013.\\\vspace{1em}
This article will appear in \textsl{IET Signal Processing}.}

\maketitle

\begin{abstract}
The newly proposed $l_1$ norm constraint zero-point attraction Least
Mean Square algorithm (ZA-LMS) demonstrates excellent performance on
exact sparse system identification. However, ZA-LMS has less
advantage against standard LMS when the system is near sparse. Thus,
in this paper, firstly the near sparse system modeling by
Generalized Gaussian Distribution is recommended, where the sparsity
is defined accordingly. Secondly, two modifications to the ZA-LMS
algorithm have been made. The $l_1$ norm penalty is replaced by a
partial $l_1$ norm in the cost function, enhancing robustness
without increasing the computational complexity. Moreover, the
zero-point attraction item is weighted by the magnitude of
estimation error which adjusts the zero-point attraction force
dynamically. By combining the two improvements, Dynamic Windowing
ZA-LMS (DWZA-LMS) algorithm is further proposed, which shows better
performance on near sparse system identification. In addition, the
mean square performance of DWZA-LMS algorithm is analyzed. Finally,
computer simulations demonstrate the effectiveness of the proposed
algorithm and verify the result of theoretical analysis.


\end{abstract}

\section{Introduction}
%
%
%
%

\begin{figure}
\centering
\includegraphics[width=4in]{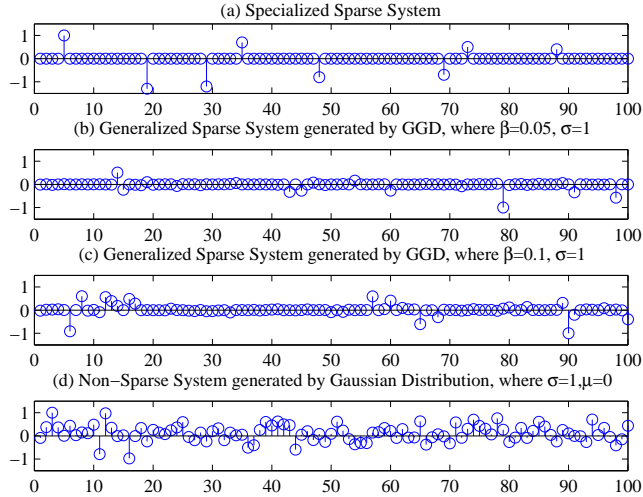}
\caption{Sparse and non-sparse systems. (a) is a exact sparse system, (b) and (c) are near sparse system generated by Generalized Gaussian Distribution with
$\beta=0.05$ and $0.1$ respectively. (d) is a non-sparse system generated by Gaussian distribution. ~} \label{sparsec}
\end{figure}

A sparse system is defined when impulse response contains only a
small fraction of large coefficients compared to its ambient
dimension. Sparse systems widely exist in many applications, such as
Digital TV transmission channel \cite{Schreiber} and the echo path
\cite{Duttweiler}. Generally, they can be further classified into
two categories: exact sparse system (ESS) and near sparse system
(NSS). If most coefficients of the impulse response are exactly
zero, it is defined as an exact sparse system
(Fig.~\ref{sparsec}.~a); Instead, if most of the coefficients are
close (not equal) to zero, it is a near sparse system
(Fig.~\ref{sparsec}.~b and c). Otherwise, a system is non-sparse if its most taps have large values (Fig.~\ref{sparsec}. d). For the simplicity of theoretical
analysis, sparse systems are usually simplified into exact sparse.
However, in real applications most systems are near sparse due to
the ineradicable white noise. Therefore, it is necessary to
investigate on near sparse system modeling and identification.

Among many adaptive filtering algorithms for system identification,
Least Mean Square (LMS) algorithm \cite{ASP}, which was proposed by
Widrow and Hoff in 60s of the past century, is the most attractive
one for its simplicity, robustness and low computation cost.
However, without utilizing the sparse characteristic, it shows no
advantage on sparse system identification. In the past few decades,
some modified LMS algorithms for sparse systems are proposed. M-Max
Normalized LMS (MMax-NLMS) \cite{Abadi} and Sequential Partial
Update LMS (S-LMS) \cite{Godavarti} reduces the computational
complexity and steady-state misalignment by partially updating the
filter coefficients. Proportionate LMS (PLMS) \cite{Duttweiler} and
its improved ones such as IPNLMS\cite{IPNLMS} and IIPNLMS\cite{IIPNLMS} accelerate the convergence
rate by updating each coefficient iteratively with different step
size proportional to the magnitude of filter coefficient. Stochastic
Tap-Normalized LMS (ST-NLMS) \cite{Y.Gu, Li} improves the
performance on specific sparse system identification where large
coefficients appear in clusters. It locates and tracks the non-zero
coefficients by adjusting the filter length dynamically. However,
its convergence performance largely depends on the span of clusters.
If the span is too long or the system has multiple clusters, it
shows no advantage compared with standard LMS algorithm.

More recently, inspired by the research of CS reconstruction problem
\cite{Donoho, J.Jin}, a class of novel adaptive algorithms for
sparse system identification have emerged based on the $l_p$ ($0\le
p\le1$) norm constraint \cite{Yi.Chen, YuantaoGu, Guolong}. Especially,
Zero-point Attraction LMS (ZA-LMS) algorithm \cite{Yi.Chen}
significantly improves the performance on exact sparse system
identification by introducing a $l_1$ norm constraint on the cost
function of standard LMS, which exerts the same zero-point
attraction force on all coefficients. However, for near sparse
systems identification, the zero-point attractor can be a
double-edged sword. Though it increases the convergence rate because
by the $l_1$ norm constraint, it also produces larger steady-state
misalignment as it forces all coefficients to exact zero. Thus, it
possesses less advantage against standard LMS algorithm when the
system is near sparse. In this paper, firstly Generalized Gaussian
Distribution (GGD)\cite{GGD} is introduced to model the near sparse
system. Then two improvements on the ZA-LMS algorithm is proposed.
Above all, by adding a window on the $l_1$ norm constraint, the
steady-state misalignment is reduced without increasing the
computational complexity. Furthermore, the zero-point attractor is
weighted to adjust the zero-point attraction by utilizing the
estimation error. By combining the two improvements, the Dynamic
Windowing ZA-LMS (DWZA-LMS) algorithm is proposed which shows
improved performance on the near sparse system identification.

The rest of the paper is organized as follows: In Section II, ZA-LMS
algorithm based on $l_1$ norm constraint is reviewed, and the near
sparse system is modeled. The new algorithm is proposed in Section
III. In Section IV, the mean square convergence performance of
DWZA-LMS is analyzed. The performances of the new algorithm and
other improved LMS algorithms for sparse system identification are
compared by simulation in Section V, where the effectiveness of our
analysis is verified as well. Finally, Section VI concludes the
paper.

\section{Preliminary}
\subsection{Review of ZA-LMS algorithm}

Let $d(n)$ be a sample of the desired output signal
\begin{equation}
d(n) = {\bf h}^{\rm T} {\bf x}(n) + v(n), \label{1}
\end{equation}
where ${\bf h} = \left[h_0,h_1,\cdots,h_{L-1}\right]^{\rm T}$ is the
unknown system with memory length $L$, ${\bf x}(n) =
[x(n),x(n-1),\cdots,x(n-L+1)]^{\rm T}$ denotes the input vector, and
$v(n)$ is the observation noise assumed to be independent of $x(n)$.
The estimation error $e(n)$ between desired and output signal is
defined as
\begin{equation}
e(n) = d(n) - {\bf w}^{\rm T}(n){\bf x}(n).\label{3}
\end{equation}
where ${\bf w}(n)$ are the filter coefficients and ${\bf w}(n) =
\left[w_0(n),w_1(n),\cdots,w_{L-1}(n)\right]^{\rm T}$. Thus the cost
function of ZA-LMS is
 \begin{equation}
 \xi_{\rm ZA}(n)=\frac{1}{2}\left|e(n)\right|^2+\eta\|{\bf w}(n)\|_1, \label{za cost function}
 \end{equation}
where $\|{\bf w}(n)\|_1$ denotes the $l_1$ norm of the filter
coefficients. Parameter $\eta$ is the factor balancing the new
penalty and the estimation error. By minimizing (\ref{za cost
function}), the ZA-LMS algorithm updates its coefficients by
\begin{equation}
{\bf w}(n+1) = {\bf w}(n)+\mu e(n) {\bf x}(n) - \rho {\rm
sgn}\left[{\bf w}(n)\right],\label{2}
\end{equation}
where $\mu$ is the step size, $\rho=\eta\mu$ is the zero-point
attraction controller, and sgn[$\cdot$] is the component-wise sign
function.

By observing (\ref{2}), the recursion of filter coefficients for
sparse system can be summarized as
\begin{equation}
{\bf w}_{\rm new} = {\bf w}_{\rm prev} + \mu f_{\rm GC} ({\bf w})+
 \rho f_{\rm ZA}({\bf w}),
\end{equation}
where $f_{\rm GC} ({\bf w}) = e(n){\bf x}$ denotes the gradient
correction function, and $f_{\rm ZA}({\bf w}) = -{\rm sgn}({\bf w})$
stands for zero-point attractor which is caused by the $l_1$ norm
penalty. For each iteration, the zero-point attractor forces the
filter taps to decrease a little when it is positive, or otherwise
to increase a little when it is negative.

\subsection{Near Sparse System Modeling}

Exact sparse system is appropriate for theoretical analysis, however,  most physical systems are near sparse with widely existing white noise in real life, thus their modeling is of more significant importance. Generalized Gaussian Distribution (GGD) is one of the
most prominent and widely used sparse distributions. For example, in
multimedia communications such as image and speech coding, GGD is
usually found to best fit the coefficients of the discrete sine and
cosine transforms, the Walsh-Hadamard transform, and the wavelet
transform \cite{Mallat}. In ultra wide bandwidth (UWB) systems, it
has recently been found to fit the multiuser interference better
\cite{Y.Chen}. These findings lead to applications of GGD in video
coding, speech recognition, blind signal separation  and UWB
receiver design \cite{K.Sharifi, H.C.Wu, S.Gazor}. Therefore GGD is
utilized to model the near sparse system in this study. It is a
class of symmetry distribution with the Gaussian and Laplacian
distribution as the special cases, with delta and uniformity
distribution as limit. The probability density function of GGD is

\begin{equation}
f(x) = \frac{\beta}{2\lambda \Gamma(1/\beta)}{\rm
exp}\left[-(|x-\mu|/\lambda)^\beta\right],\label{4}
\end{equation}
where $\lambda =
\sigma_g\sqrt{\frac{\Gamma(1/\beta)}{\Gamma(3/\beta)}}$, $\Gamma(x)$
denotes the Gamma function, $\mu$ and $\sigma_g^2$ are called the
mean and variance of GGD, respectively. Besides, $\beta$ determines
the decay rate of the density function and can be used to denote the
sparsity of the system (see as Fig.~\ref{sparsec}.~b, c; please notice that the system sparsity  decreases as $\beta$ increases).
Especially, GGD is Gaussian Distribution when $\beta=2$, and it
turns into Laplacian Distribution when $\beta=1$. By integrating
$f(x)$, the distribution function is
\begin{equation}
F(x)=\frac{1}{2}+{\rm
sgn}(x-\mu)\frac{\Theta\left[1/\beta,\left(\frac{|x-\mu|}{\lambda}\right)^\beta\right]}{2\Gamma(1/\beta)}.\label{5}
\end{equation}
where $\Theta(s,x) = \int_0^x t^{s-1}e^{-t}{\rm d}t$ is called the
lower incomplete function.

\section{Improved ZA-LMS Algorithm}

\begin{figure}[!t]
\centering
\includegraphics[width=4in]{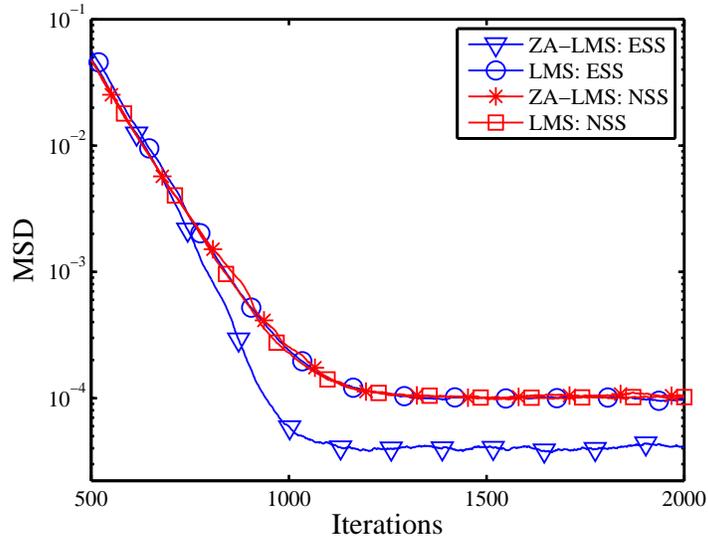}
\caption{Comparison between ZA-LMS and LMS on Exact Sparse System and Near Sparse System.} \label{Drawback}
\end{figure}

The ZA-LMS algorithm improves the performance by exerting an $l_1$
norm penalty forcing small coefficients to zero iteratively. It is
very effective for the exact sparse system identification. However,
for the near sparse system with small white noise on all
coefficients, most of the small coefficients are forced to zero. Thus
ZA-LMS usually degrades with large steady-state misalignment,
showing no improvement compared with standard LMS. The situation is
more clearly illustrated in Fig.~\ref{Drawback}, where the exact
sparse sparse system is generated with 8 nonzero large coefficients
with tap length of 100, and the near sparse system is the same with
the former except that a power of $1\times10^{-4}$ white noise is
added on all coefficients. The step sizes for all algorithms are set
as $\mu=0.01$. The optimal parameter $\rho$ is derived theoretically
from (36) of \cite{K.Shi} by minimizing the steady-state MSD.

From Fig.~\ref{Drawback}, it can be seen that by choosing the
theoretically optimal parameter $\rho$ the performance of ZA-LMS is
much better than standard LMS for exact sparse system. However, the
performance of ZA-LMS severely degrades on near sparse system
identification. As stated above, the main reason is that the strong
zero-point attraction forces near zero coefficients to zero that
results in the large steady-state misalignment. Though empirically
decreasing $\rho$ may alleviate this problem, in that case, ZA-LMS
will degrade to standard LMS showing no improvement, as shown in
Fig.~\ref{Drawback}. Thus to improve the performance of near sparse
system identification, a new ZA-LMS based algorithm is put forward,
the main modifications are as follows.

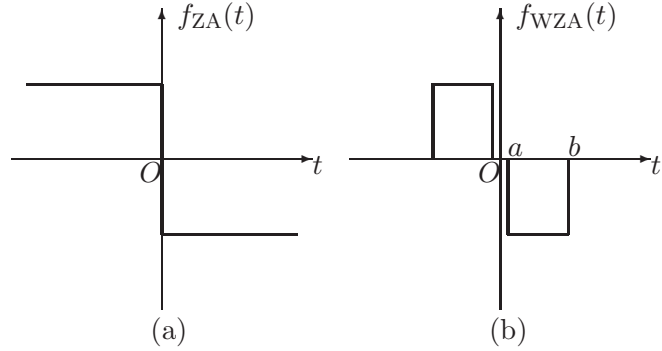
\begin{figure}\label{zaterm}
\begin{center}
\setlength{\unitlength}{1mm}
\begin{picture}(90,44)
\thinlines
\put(0,24){\vector(1,0){40}}
\put(20,4){\vector(0,1){40}}
\thicklines
\put(40,22){$t$}
\put(17,21){$O$}
\put(22,42){$f_{\rm ZA}(t)$}
\put(18.5,0){(a)}
\put(20,14){\line(1,0){18}}
\put(20,34){\line(-1,0){18}}
\put(20,14){\line(0,1){20}}
\thinlines
\put(45,24){\vector(1,0){40}}
\put(65,4){\vector(0,1){40}}
\thicklines
\put(85,22){$t$}
\put(67,42){$f_{\rm WZA}(t)$}
\put(62,21){$O$}
\put(66,24.5){$a$}
\put(74,24.5){$b$}
\put(63.5,0){(b)}
\put(66,14){\line(1,0){8}}
\put(66,14){\line(0,1){10}}
\put(74,14){\line(0,1){10}}
\put(64,34){\line(-1,0){8}}
\put(64,34){\line(0,-1){10}}
\put(56,34){\line(0,-1){10}}
\end{picture}
\end{center}
\caption{The zero-point attractor for sparse system, (a)ZA-LMS, (b)WZA-LMS.}
\end{figure}

\subsection{Windowing ZA-LMS}
As the attracting range of ZA-LMS algorithm reaches infinity, all
coefficients of the sparse system are attracted to zero point. However, the
identical attraction on both large and small ones will lead to
increase the computational complexity and steady-state misalignment.
Thus, the first improvement lies in the constraint of zero-point
attracting range. As shown in Fig.~3, the new zero-point
attraction, which attracts the coefficients only in a certain range,
is proposed by adding a window on the original zero-point
attracting. The new recursion of coefficients of the proposed
Windowing ZA-LMS (WZA-LMS) is
 \begin{equation}
{\bf w}(n+1) = {\bf w}(n)+\mu e(n) {\bf x}(n) - \rho {\rm
sgn}_w\left[{\bf w}(n)\right],\label{wzalms}
\end{equation}
where ${\rm sgn}_w[\cdot]$ is the component-wise partial sign
function, defined as
\begin{equation}
{\rm sgn}_w[t]=-f_{\rm WZA}(t)=\left\{
\begin{array}{ll}
{\rm sgn}(t) &\mbox{$a<|t|\le b$;}\\
0 &\mbox{elsewhere.}\label{new sgn}
\end{array}
\right.
\end{equation}
where $a$ and $b$ are both positive constant which denotes the lower
and upper threshold of the attraction range, respectively.

From Fig.~3, it can be concluded that ZA-LMS is the
special case of WZA-LMS when $a$ reaches 0 and $b$ approaches
infinity, respectively. Besides, by investigating (\ref{2}),
(\ref{wzalms}) and (\ref{new sgn}), it can be seen that the
computational complexity of the two algorithms is approximately the
same. And by adopting the new zero-point attractor and properly
setting the threshold, the coefficients, whether too small or too
large, will not be attracted any more. Thus, the steady-state
misalignment is significant reduced especially for near sparse
system.

\subsection{Dynamic ZA-LMS}
As mentioned above, the sparse constraint should be relaxed in order
to reduce the steady-state misalignment when the updating procedure
reaches the steady-state. Inspired by the idea of variable step size
methods of standard LMS algorithm\cite{Kwong}, the magnitude of
estimation error, which denotes the depth of convergence, is
introduced here to adjust the force of zero-point attraction
dynamically. That is,
\begin{equation}\label{fnew}
    f_{\rm DZA}(t) = f_{\rm WZA}(t)|e(n)|.
\end{equation}
At the beginning of iterations, large estimation error increases
zero-point attraction force which also accelerates the convergence.
When the algorithm is approaching the steady-state, the error
decreases to a minor value accordingly. Thus the influence of
zero-point attraction force on small coefficients is reduced that
produce smaller steady-state misalignment. By implementation of this
improvement on ZA-LMS algorithm, the algorithm is named as Dynamic
ZA-LMS (DZA-LMS).

\subsection{Dynamic Windowing ZA-LMS}
Finally, by combining the two improvements, the final Dynamic
Windowing ZA-LMS (DWZA-LMS) algorithm can be drew. The new recursion
of filter coefficients is as follows,
\begin{eqnarray}
w_i(n+1) = w_i(n)+\mu e(n)x(n-i)-\rho|e(n)|{\rm
sgn}_w[w_i(n)]\quad\forall 0\leq i < L.\label{6}
\end{eqnarray}

In addition, the new method can also improve the performance of
ZA-NLMS, which is known for its robustness. The  recursion of
DWZA-NLMS is
\begin{equation}
w_i(n+1) = w_i(n)+\frac{1}{\epsilon+{\bf x}^{\rm T}(n){\bf
x}(n)}\left\{\mu e(n)x(n-i)-\rho\left|e(n)\right|{\rm
sgn}_w\left[w_i(n)\right]\right\} \quad\forall 0\leq i < L.\label{7}
\end{equation}
where $\epsilon>0$ is the regularization parameter.

\section{Analysis of the proposed Algorithm}

The mean square convergence analysis of DWZA-LMS algorithm is
carried out in this section. The analysis is based on the following
assumptions.
\begin{enumerate}
  \item The input signal $x(n)$ is i.i.d zero-mean Gaussian. The observation noise $v(n)$ is zero-mean white. The tap-input vectors ${{\bf x}(n)}$ and the desired response $d(n)$ follow the common independence assumption \cite{Haykin}, which are generally used for performance analysis of LMS
algorithm.
  \item The unknown near sparse filter tap follows GGD. As stated in Section II, this assumption is made because GGD is the suitable sparse distribution for near sparse system
modeling.
  \item The steady state adaptive filter tap $w_i(n)$ follows the same distribution with $h_i$ ($1\le i \le L$). This is a reasonable assumption in that the error between the coefficients of the identified and the unknown real systems are very small when the algorithm converges.

\end{enumerate}

Under these assumptions, the mean square convergence condition and
the steady-state MSE of DWZA-LMS algorithm are derived. Also, the
choice of parameters is discussed in the end of this section.

First of all, the misalignment vector is defined as
\begin{equation}
{\bm \Delta}(n) = {\bf w}(n) - {\bf h},\label{8}
\end{equation}
and auto-covariance matrix of ${\bf \Delta}(n)$ as
\begin{equation}
{\bf K}(n)={\rm E}\left\{{\bf \Delta}(n){\bm \Delta}^{\rm
T}(n)\right\}.\label{9}
\end{equation}

\subsection{Mean Square Convergence of Misalignment}

Combining (\ref{1}), (\ref{3}), (\ref{6}) and (\ref{8}), one
derives
\begin{equation}
{\bm \Delta}(n+1) = {\bf A}(n){\bm \Delta}(n) +\mu v(n){\bf x}(n)-
\rho {\bf m}(n),\label{10}
\end{equation}
where ${\bf A}(n) = {\bf I}-\mu{\bf x}(n){{\bf x}^{\rm T}(n)}$,
${\bf m}(n)=\left[m_0(n),m_1(n),\cdots,m_{L-1}(n)\right]$, and
\begin{equation}
m_i(n) = |e(n)|{\rm sgn}_w[w_i(n)] \quad\forall 0\leq i <
L.\label{11}
\end{equation}
By utilizing the independence assumption\cite{ASP}, and substituting
(\ref{10}) into (\ref{9}) yields
\begin{eqnarray}
{\bf K}(n+1)&=&{\rm E}\{{\bf A}(n){\bm \Delta}(n){\bm \Delta}^{\rm T}(n){\bf A}^{\rm T}(n)\}+\mu^2\sigma_v^2\sigma_x^2{\bf I}_L\nonumber\\
&&-\rho{\rm E}\{{\bf m}(n){\bm \Delta}^{\rm T}(n){\bf A}^{\rm T}(n)\}+\rho^2{\rm E}\{{\bf m}(n){\bf m}^{\rm T}(n)\}\nonumber\\
&&-\rho{\rm E}\{{\bf A}(n){\bm \Delta}(n){\bf m}^{\rm
T}(n)\},\label{12}
\end{eqnarray}
where ${\bf I}_L$ is an $L\times L$ unit matrix, $\sigma_x^2$ and
$\sigma_v^2$ denote the power of input signal and observation noise,
respectively. By utilizing the property that the fourth-order moment
of a Gaussian variable is three times the variance square, one
obtains
\begin{equation}
{\rm E}\{{\bf A}(n){\bm \Delta}(n){\bm \Delta}^{\rm T}(n){\bf
A}^{\rm T}(n)\} = (1-2\mu\sigma_x^2+2\mu^2\sigma_x^4){\bf K}(n)
+\mu^2\sigma_x^4D(n){\bf I}_L,\label{13}
\end{equation}
where $D(n)={\rm tr}\left[{\bf K}(n)\right]$. With (\ref{8}), one
has
\begin{eqnarray}
{\rm E}\{{\bf A}(n){\bm \Delta}(n){\bf m}^{\rm T}(n)\} = \left\{{\rm
E}\{{\bf m}(n){\bm \Delta}^{\rm T}(n){\bf A}^{\rm
T}(n)\}\right\}^{\rm T} =(1-\mu\sigma_x^2){\rm E}\{{\bm
\Delta}(n){\bf m}^{\rm T}(n)\}.\label{14}
\end{eqnarray}
Combining (\ref{12}), (\ref{13}) and (\ref{14}), one derives
\begin{eqnarray}
{\bf K}(n+1) &=& (1-2\mu\sigma_x^2+2\mu^2\sigma_x^4){\bf K}(n)+\mu^2\sigma_x^4D(n){\bf I}_L\nonumber\\
&&+\mu^2\sigma_v^2\sigma_x^2{\bf I}_L-\rho(1-\mu\sigma_x^2)\left\{{\rm E}\{{\bm \Delta}(n){\bf m}^{\rm T}(n)\}\right\}^{\rm T}\nonumber\\
&&-\rho(1-\mu\sigma_x^2){\rm E}\{{\bm \Delta}(n){\bf m}^{\rm
T}(n)\}+\rho^2{\rm E}\{{\bf m}(n){\bf m}^{\rm T}(n)\}.\label{15}
\end{eqnarray}
By taking trace on both sides of (\ref{15}), it can be concluded
that the adaptive filter is stable if and only if
\begin{equation}
0<1-2\mu\sigma_x^2+(L+2)\mu^2\sigma_x^4<1,\label{16}
\end{equation}
which is simplified to
\begin{equation}
0<\mu<\frac{2}{(L+2)\sigma_x^2}.\label{17}
\end{equation}

This implies that the proposed DWZA-LMS algorithm has the same
stability condition for the mean square convergence as the ZA-LMS
and standard LMS algorithm\cite{ASP}.

\subsection{Steady-state Mean Square Error}
In this subsection, the steady-state Mean Square Error (MSE) of
DWZA-LMS algorithm is analyzed. By definition, MSE is
\begin{equation}
{\rm MSE}={\rm E}\{e^2(\infty)\}
   =\sigma_v^2+{\rm tr}\{{\bf R}(\infty){\bf K}(\infty)\},\label{18}
\end{equation}
where ${\bf R}(\infty)=\sigma_x^2{\bf I}_L$, then (\ref{18}) can be
rewritten as
\begin{equation}
{\rm MSE}=\sigma_v^2+\sigma_x^2D(\infty).\label{19}
\end{equation}
Thus, our work is to estimate $D(\infty)$. In (\ref{15}), let $n$
approach infinity, by observing the $i$th ($0\le i<L$) element
$K_i(\infty)$ of the matrix ${\bf K}(\infty)$, one obtains
\begin{eqnarray}
K_i(\infty)&=&(1-2\mu\sigma_x^2+2\mu^2\sigma_x^4)K_i(\infty)+\mu^2\sigma_x^4D(\infty)\nonumber\\
&&+\mu^2\sigma_v^2\sigma_x^2-2\rho(1-\mu\sigma_x^2){\rm E}\{\Delta_im_i(\infty)\}\nonumber\\
&&+\rho^2{\rm E}\{m_i^2(\infty)\}\quad\forall 0\leq i < L,\label{20}
\end{eqnarray}
With reference to (\ref{11}), it is obvious that
\begin{equation}
{\rm E}\{m_i^2(\infty)\}=\left\{
\begin{array}{ll}
{\rm MSE} &\mbox{$a<|w_i(\infty)|\le b$;}\\
0 &\mbox{elsewhere.}\label{21}
\end{array}
\right.
\end{equation}
To derive ${\rm E}\{\Delta_i(\infty)m_i(\infty)\}$, by multiplying
${\bf m}^{\rm T}(n)$ on the right of each item of (\ref{10}) and
taking the expectation value on both sides as well as letting $n$
approach infinity it yields
\begin{equation}
{\rm E}\{{\bm \Delta}(\infty){\bf m}^{\rm T}(\infty)\} =
-\frac{\rho}{\mu\sigma_x^2}{\rm E}\{{\bf m}(\infty){\bf m}^{\rm
T}(\infty)\}.\label{22}
\end{equation}
Thus, when $|w_i(\infty)|\le a$ or $|w_i(\infty)|>b$ ($0\le i<L$),
it has
\begin{equation}
K_i(\infty)=\frac{\mu\sigma_x^2D(\infty)}{2(1-\mu\sigma_x^2)}+\frac{\mu^2\sigma_x^2\sigma_v^2}{2\mu\sigma_x^2(1-\mu\sigma_x^2)},\label{23}
\end{equation}
when $a<|w_i(\infty)|\le b$ ($0\le i<L$), it has
\begin{equation}
K_i(\infty) =
\frac{1}{2\mu\sigma_x^2(1-\mu\sigma_x^2)}\left[\mu^2\sigma_x^4D(\infty)+\mu^2\sigma_v^2\sigma_x^2
+\left(\frac{2}{\mu\sigma_x^2}-1\right)\rho^2{\rm
MSE}\right].\label{24}
\end{equation}

According to Assumption (3), and combining (\ref{5}), we have
\begin{eqnarray}
{\rm P}_A&=&{\mathcal P}\left\{a \le |w_i(n)|\le b\right\}\nonumber\\
&=&2\left[F(b)-F(a)\right]\nonumber\\
&=&\left\{\Theta\left[1/\beta,(\frac{|b|}{\lambda})^\beta\right]-\Theta\left[1/\beta,(\frac{|a|}{\lambda})^\beta\right]\right\}
/\Gamma(1/\beta),
\end{eqnarray}
where ${\rm P}_A$ denotes the probability that the coefficients of
adaptive filter will be attracted. On the other hand, the
probability that they will not be attracted is ${\rm P}_{NA}=1-{\rm
P}_A$. By combining (\ref{23}) and (\ref{24}) and summing up all the diagonal items of
matrix ${\bf K}(n)$, it yields
\begin{equation}
D(\infty)=\frac{L\left[\mu^2\sigma_v^2\sigma_x^2+\rho^2{\rm
P}_A\left({\displaystyle\frac{2}{\mu\sigma_x^2}}-1\right){\rm
MSE}\right]}{\mu\sigma_x^2\left[2-\mu\sigma_x^2(L+2)\right]},\label{25}
\end{equation}
Combining (\ref{19}) and (\ref{25}) , finally one has
\begin{equation}
{\rm
MSE}=\frac{\mu\sigma_v^2\left[2-\mu\sigma_x^2(L+2)\right]+L\mu^2\sigma_x^2\sigma_v^2}{2\mu-\mu^2\sigma_x^2(L+2)-\rho^2{\rm
P}_AL\left({\displaystyle\frac{2}{\mu\sigma_x^2}}-1\right)}.\label{MSE}
\end{equation}
If $\rho=0$, equation (\ref{MSE}) is the same with MSE of standard
LMS algorithm \cite{ASP},
\begin{equation}
{\rm
MSE}_{\rm{LMS}}=\frac{\left(2-2\mu\sigma_x^2\right)\sigma_v^2}{2-\mu\sigma_x^2(L+2)}.\label{MSELMS}
\end{equation}

\subsection{Parameter Analysis}
The performance of the proposed algorithm is largely affected by the
balancing parameter $\rho$ and the thresholds $a$ and $b$.

According to (\ref{za cost function}) and (\ref{2}), it can be seen
that the parameter $\rho$ determines the importance of the $l_1$
norm and the intensity of zero-point attraction. In a certain range,
a larger $\rho$, which indicates stronger attraction intensity, will
improve the convergence performance by forcing small coefficients
toward zero with fewer iterations. However, according to
(\ref{MSE}), a larger $\rho$ also results in a larger steady-state
misalignment. So the parameter $\rho$ can balance the tradeoff
between adaptation speed and quality. Moreover, the optimal
parameter $\rho$ empirically satisfies $\rho\ll\mu\ll1$. By
analyzing steady-state MSE in (\ref{MSE}) under such circumstance,
it can be seen that
\begin{equation}\label{compare}
\left|\rho^2{\rm
P}_AL\left(\frac{2}{\mu\sigma_x^2}-1\right)\right|\ll\left|2\mu-\mu^2\sigma_x^2(L+2)\right|.
\end{equation}
According to (\ref{compare}), the influence of the last term in the
denominator of (\ref{MSE}) can be ignored, which means that the
steady-state MSE of the proposed algorithm is approximately the same
with standard LMS for near sparse systems identification. The same
conclusion can also be drawn from (\ref{6}) intuitively: when the
adaptation reaches steady-state, the small $e(n)$ renders the value
of $\rho|e(n)|$ trivial compared to $\mu$, letting the relaxation of
zero-point attraction constraint. On the other hand, with large
$e(n)$ in the beginning and the process of adaptation which
indicates larger zero-point attraction force, the zero-point
attractor adjusts the small taps more effectively than ZA-LMS,
forcing them to zero with fewer iterations, which accelerate the
convergence rate significantly.

The thresholds $a$ and $b$ determine the zero-point attraction range
together. The parameter $a$ is set to avoid forcing all small coefficients to exact zero, it is suggested to be set as the mean amplitude of those near zero coefficients of the real system. Specifically, for exact sparse systems, as most coefficients of the
unknown system are exactly zero except some large ones, accordingly
$a=0$ is set to force most small coefficients to exact zero. For exact sparse systems contaminated by small Gaussian white noise, $a$ should be set as the standard deviation of the noise. For near sparse systems generated by GGD, as the mean amplitude of the small coefficient is hard to derive, we empirically choose $a$ for the proposed algorithm.  As a small sparsity indicator $\beta$ in GGD usually means smaller mean amplitude of the small coefficient, we choose smaller $a$ when $\beta$ is smaller. According to the simulations,  $a$ is chosen in the range $1\times 10^{-3}$ to $1 \times 10^{-2}$ for GGD with $\beta$ varying from $0.05$ to $0.5$. The parameter $b$ is chosen to reduce the unnecessary attraction of large coefficients in ZA-LMS, therefore, empirically any constant $b$, which is much larger than the deviation of small coefficients and much smaller than infinity, should be appropriate. Various simulations demonstrate that the parameter $b$ can be set as a constant around 1 for most near sparse systems. This choice of $b$ is quite standard for most applications.

\section{Simulations}

In this section, first we demonstrate the convergence performance of our proposed algorithm on two near sparse systems and a exact sparse system in Experiment 1-4, respectively. Second, Experiment 5-7 are designed to verify the derivation and discussion in Section IV. Besides the proposed algorithm, Standard NLMS, ZA-LMS, IPNLMS\cite{IPNLMS} and IIPNLMS\cite{IIPNLMS} are also simulated for comparison. To be noticed, the normalized variants of ZA-LMS and the proposed algorithm are adopted to guarantee a fair comparison in all experiments except the fourth, where DWZA-LMS is simulated to verify the theoretical analysis result.

The first experiment is to test the convergence and tracking
performance of the proposed algorithm on near sparse
system driven by Gaussian white signal and correlated input, respectively. The unknown system is generated by GGD which has been shown in Fig.~\ref{sparsec}.~b with filter
length $L=100$, it is initialized randomly with $\beta=0.05$ and
$\sigma_g^2=1$. For the white and correlated input, the system is regenerated following the same
distribution after $1700$ and $4500$ iterations, respectively. For the white input, the signal $x(n)$ is generated by white Gaussian noise with power
$\sigma_x^2=1$. For the correlated input, the signal is generated by white Gaussian noise $y(n)$ driving a first-order Auto-Regressive (AR) filter, $x(n)=0.8x(n-1)+y(n)$, and $x(n)$ is normalized. Besides, the power of observation noise is $\sigma_v^2=1\times10^{-4}$ for both input. The five algorithms are simulated 100 times respectively with parameter $\mu=1$ in both cases. The other parameters
are as follows
\begin{itemize}

\item IPNLMS and IIPNLMS with white input: $\alpha_{\rm P}=-0.5$, $\rho=0.4$, $\alpha_{\rm P1}=-0.5$, $\alpha_{\rm P2}=0.5$, $\Gamma=0.1$;

\item IPNLMS and IIPNLMS with correlated input: $\alpha_{\rm P}=-0.5$, $\rho=0.2$, $\alpha_{\rm P1}=-0.5$, $\alpha_{\rm P2}=0.5$, $\Gamma=0.1$;

\item ZA-NLMS and DWZA-NLMS with white input: $\rho_{\rm ZA}=3\times10^{-4}$, $\rho_{\rm DWZA}=6\times10^{-2}$, $a=1\times10^{-3}$, $b=0.8$.
\item ZA-NLMS and DWZA-NLMS with correlated input: $\rho_{\rm ZA}=3\times10^{-4}$, $\rho_{\rm DWZA}=3\times10^{-2}$, $a=1\times10^{-3}$, $b=0.8$.

\end{itemize}

\begin{figure}[!t]
\centering
\subfigure[White Input]{
\label{experiment_1_a}
\includegraphics[width=4in]{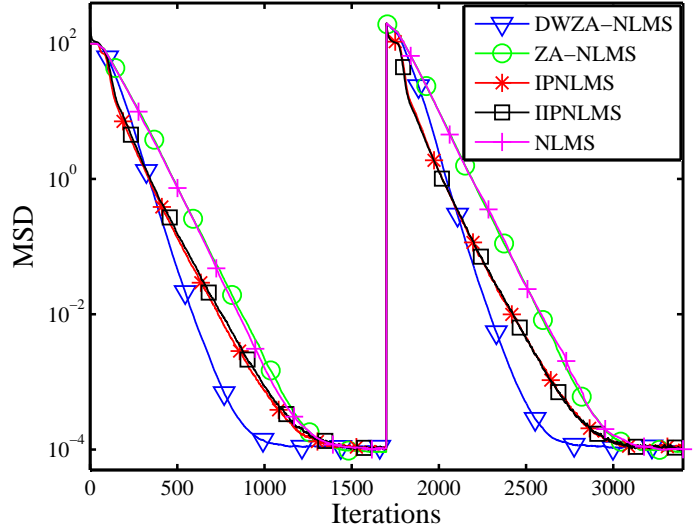}}
\subfigure[Correlated Input]{
\label{experiment_1_b}
\includegraphics[width=4in]{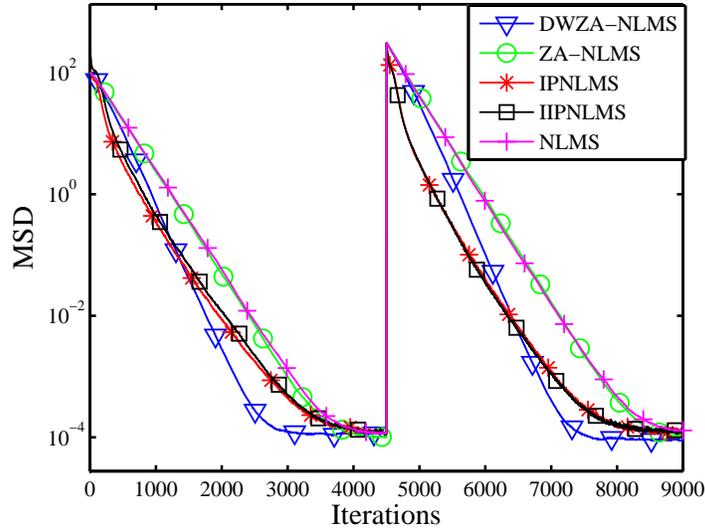}}
\caption{Comparisons of convergence rate and tracking ability
 of five different algorithms for sparse system identification driven by (a) white
input and (b) correlated input, respectively. }
\label{experiment_1}
\end{figure}

\begin{table}[t]
\renewcommand{\arraystretch}{1.3}
\caption{Comparison of computational complexity of IPNLMS, IIPNLMS
and DWZA-NLMS} \label{computational complexity}

\begin{center}\begin{tabular}{ccccc}
\hline
{\raisebox{-2ex}[0pt][0pt]{Algorithms}} &    &     Multiplies         &                     &\raisebox{-2ex}[0pt][0pt]{Comparisons}\\
\cline{2-4}
           &Convolution   &Tap update    &Total\\
\hline
IPNLMS     &$L$             &$3L$            &$4L$             &$0$\\
IIPNLMS    &$L$             &$3L$            &$4L$             &$4L$\\
ZA-NLMS &$L$             &$2L$            &$3L$             &$0$\\
DWZA-NLMS   &$L$             &$(1+P_1)L$      &$(2+P_1)L$         &$2L$\\
\hline
\end{tabular}\end{center}
Where $P_1\in [0,1] $ denotes the fraction of coefficients in the attracting range of DWZA-NLMS. \\
\end{table}

All the parameters are particularly selected to keep their
steady-state error in the same level. The MSD of these algorithms for both white and correlated input
are shown in Fig.~\ref{experiment_1_a} and Fig.~\ref{experiment_1_b}, respectively. The simulation results show that all algorithms converge more slowly in the color input driven scenario than in the white noise driven case. However, the ranks or their relative performances are similar and the proposed algorithm reaches the steady state first with both white and correlated input. On the other hand, the performance of ZA-NLMS degenerate to standard NLMS as the system is near sparse. Furthermore, the computational complexity of the proposed algorithm is also smaller compared with improved PNLMS algorithms (Table \ref{computational complexity}). Besides, when the system is changed abruptly, the proposed algorithm also reaches the steady-state first in both cases.

\begin{figure}[!t]
\centering
\includegraphics[width=4in]{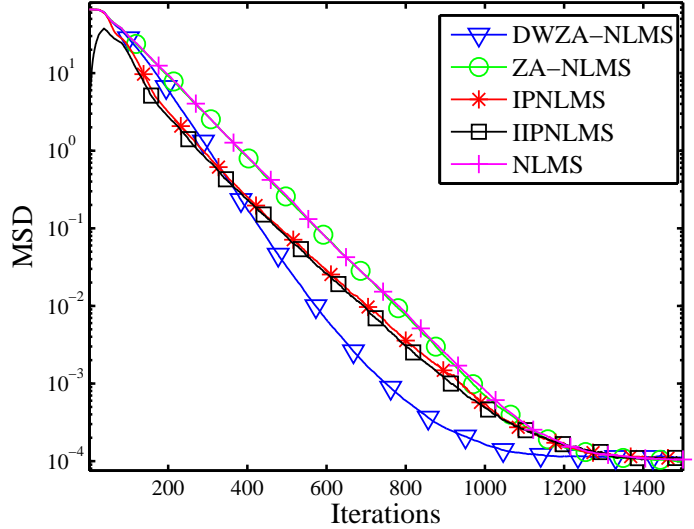}
\caption{Comparisons of convergence rate on NLMS, ZA-NLMS, IPNLMS, IIPNLMS, and the proposed algorithm on near sparse system.} \label{experiment_5}
\end{figure}

The second experiment is to demonstrate the proposed algorithm on near-sparse systems other than GGD. The near-sparse system with 100 taps is generated in the following manner. First, 8 large coefficients following Gaussian distribution $\mathcal{N}(0,1)$ are generated, where all their tap positions follow Uniform distribution. Second, white Gaussian noise with variance $\sigma_h^2 = 1\times 10^{-4}$ is added to all taps, enforcing the system to be near-sparse. The signal $x(n)$ is generated by white Gaussian noise with power $\sigma_x^2=1$. Five algorithms, the same as in Experiment 1, are simulated 100 times respectively with parameter $\mu=1$. The parameters are set to the same values as in the white input case of Experiment 1 except $a = 1\times10^{-2}$ for the proposed algorithm. From Fig. \ref{experiment_5}, we can conclude that the proposed algorithm reaches the steady-state first in such near sparse system.

The third experiment demonstrates the effectiveness of the proposed modification on ZA-LMS algorithm on near sparse system. The system and the signal are generated in the same way as in Experiment 2. The proposed DZA-NLMS, DWZA-NLMS are compared with ZA-NLMS for the system with 100 simulations. The step length for all algorithms are set as $\mu=1$. We particularly choose parameters $\rho_{DZA}=0.05$ and $\rho_{ZA}= 6.5\times 10^{-4}$ for  DZA-NLMS and ZA-NLMS to ensure their steady-state mean square error in the same level. We set $\rho_{DWZA}=0.05$ for DWZA-NLMS algorithm for a fair comparison with DZA-NLMS. The parameter $a$ and $b$ in DWZA-NLMS are chosen as $a=0.01$ and $b=0.8$, respectively.

\begin{figure}[!t]
\centering
\includegraphics[width=4in]{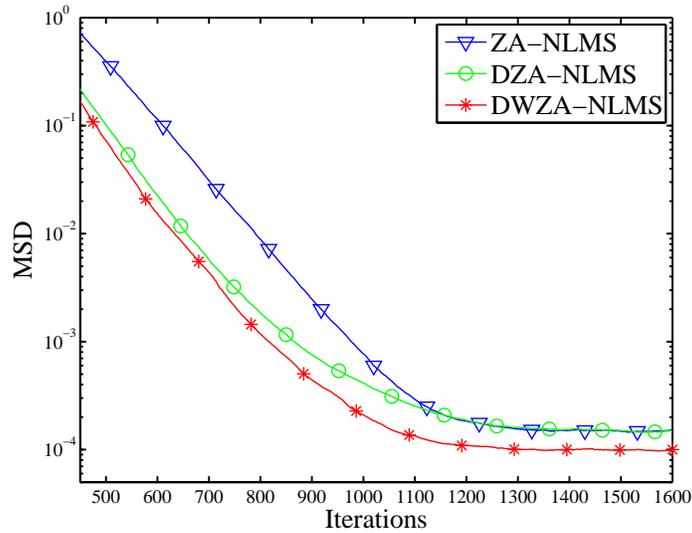}
\caption{Comparisons of convergence speed on ZA-NLMS, DZA-NLMS and DWZA-NLMS.} \label{experiment_7}
\end{figure}

From Fig. \ref{experiment_7}, we can see that with the dynamic zero-point attractor DZA-NLMS convergences faster than ZA-NLMS. By adding another window constraint on the zero-point attractor, The DWZA-NLMS not only preserves the property of fast convergence of DZA-NLMS, but also shows smaller steady-state mean square error than both ZA-NLMS and DZA-NLMS.

\begin{figure}[!t]
\centering
\includegraphics[width=4in]{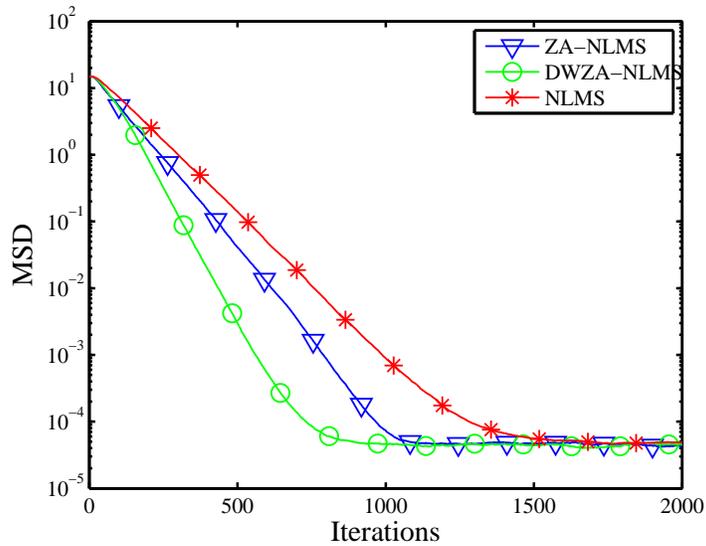}
\caption{Comparisons of convergence rate on NLMS, ZA-NLMS, and the proposed algorithm on exact sparse system.} \label{experiment_4}
\end{figure}

The fourth experiment shows that the proposed improvement
is still effective on the exact sparse system identification. The
unknown system is shown in Fig.~\ref{sparsec}.~a, where filter
length $L=100$. Besides, 8 large coefficients is uniformly distributed and generated by Gaussian distribution $\mathcal{N}(0,1)$, and all other tap coefficients are exactly zero. The input signal is generated by white Gaussian noise with power $\sigma_x^2=1$, and the power of observation noise is $\sigma_v^2=1\times10^{-4}$. The proposed
algorithm is compared with ZA-NLMS and standard NLMS, where each
algorithm is simulated 100 times with 2000 iterations. The step
size $\mu$ is set to $0.65$ for NLMS, and $\mu=1$ for both the ZA-NLMS and
the proposed algortihms. The parameter $\rho_{\textrm{ZA}}$ is set to $6\times 10^{-4}$ and $\rho_{\textrm{DWZA}}=6\times 10^{-2}$. All the parameters are chosen to make sure
that the steady-state error is the same for comparison. According to
Fig.~\ref{experiment_4}, it can be seen that the convergence
performance is also improved compared with ZA-NLMS via the proposed
method on exact sparse system, thus the proposed improvement on the
algorithms are robust.

\begin{figure}[!t]
\centering
\includegraphics[width=4in]{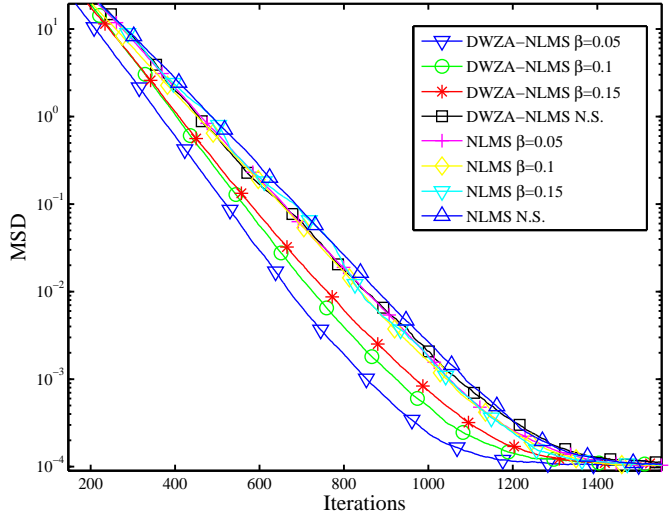}
\caption{Comparisons of convergence rate of proposed algorithm for different sparsity.} \label{experiment_3}
\end{figure}

The fifth experiment is to test the sensitivity to sparsity of the
proposed algorithm. All conditions are the same with the
first experiment except the sparsity. The parameter $\beta$ is
selected as $0.05, 0.1$, and $0.15$, respectively. Besides, we also compared our algorithm when the system is non-sparse which is generated by Gaussian distribution. For each $\beta$, both the proposed algorithm and NLMS are simulated $50$ times with $1700$
iterations. The step size $\mu$ is $1$ for both algorithms, and
$\rho=4\times10^{-2}$, $a=1\times10^{-2}$, $b=0.8$ for the proposed algorithm. The
simulated MSD curves are shown in Fig.~\ref{experiment_3}. The steady-state MSD remains approximately the same for the proposed algorithm with varying parameter $\beta$ which denotes the sparsity of systems, meanwhile the convergence rate decreases as the sparsity decreases. For the non-sparse case, our algorithm degenerates and shows similar behavior with standard NLMS. However, for each $\beta$ the proposed algorithm is never slower than standard NLMS. It should be noticed that NLMS is independent on the system sparsity and behaves similar when $\beta$ varies.

\begin{figure}[!t]
\centering
\includegraphics[width=4in]{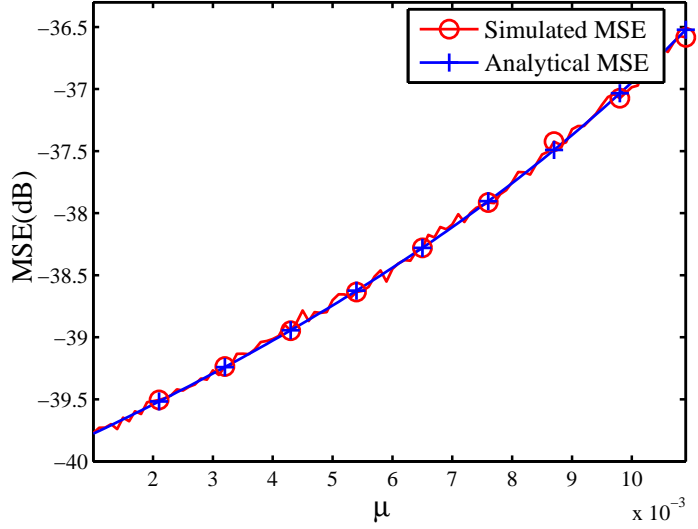}
\caption{ The steady-state MSD of the proposed algorithm with different step size} \label{experiment_2_mu}
\end{figure}

\begin{figure}[!t]
\centering
\includegraphics[width=4in]{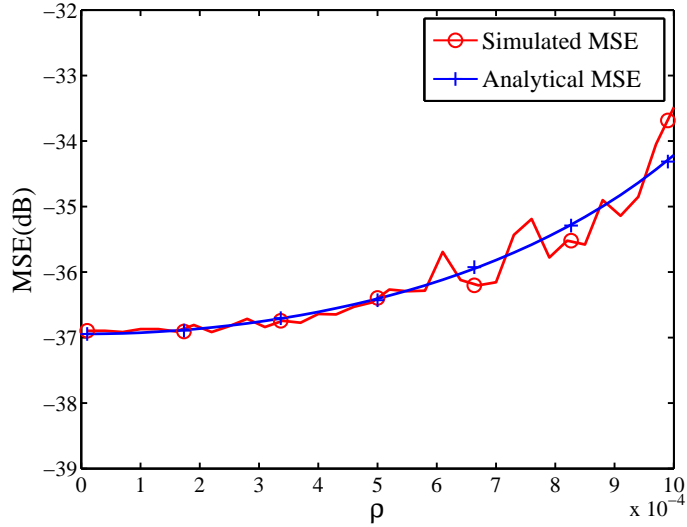}
\caption{ The steady-state MSD of the proposed algorithm algorithm with different parameters $\rho$} \label{experiment_2_rho}
\end{figure}

The sixth experiment is to test the steady-state MSE with different
parameters. The coefficients of unknown system follows GGD with
filter length $L=100$, the sparsity and variance are chosen as
$\sigma_g^2=1$ and $\beta=0.1$, respectively. The input is generated
by white Gaussian noise with normalized power $\sigma_x^2=1$, and
the power of observation noise is $1\times10^{-4}$. Under such
circumstance, the steady-state MSD is tested. Here $a=1\times10^{-2}$ and $b=0.8$ are
set for each simulation. The step size $\mu$ is varied from 0 to
$1.1\times10^{-3}$ for given $\rho=2\times10^{-4}$. And $\rho$ is
changed from 0 to $1\times10^{-3}$ for given $\mu=1\times10^{-2}$.
Fig.~\ref{experiment_2_mu} and Fig.~\ref{experiment_2_rho} show that
the analytical results accord with the simulated ones of different
parameters for variable values. Specifically, in
Fig.~\ref{experiment_2_mu}, the steady-state MSE goes up as the step
size increases, whose trend is the same with standard LMS.
Fig.~\ref{experiment_2_rho} shows that the analytical steady-state
MSE matches with simulated one as the parameter $\rho$ gets larger,
which verifies the result in Section V.

\begin{figure}[!t]
\centering
\subfigure[Fix $b=0.8$ while varying $a$ from $0$ to $0.1$.]{
\label{experiment_6_a}
\includegraphics[width=0.5\textwidth]{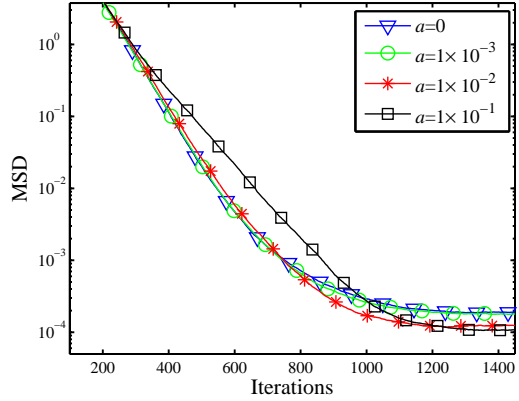}}
\subfigure[Fix $a=1\times 10^{-2}$ while varying $b$ from $0.1$ to $5$.]{
\label{experiment_6_b}
\includegraphics[width=0.5\textwidth]{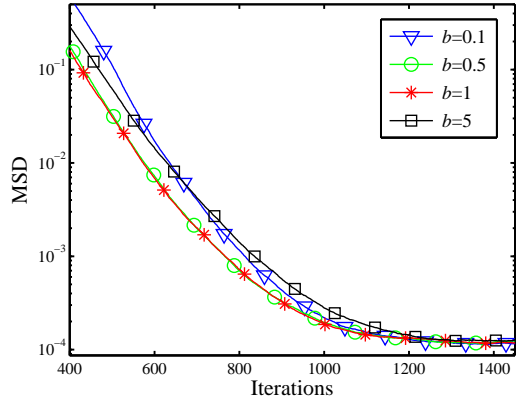}}
\caption{The effect of parameters $a$ and $b$ on the proposed algorithm for near sparse system.}
\label{experiment_6}
\end{figure}

The seventh experiment is designed to test the behavior of the proposed algorithm with respect to different parameters of $a$ and $b$. All conditions of the proposed algorithm are the same with the second experiment except the parameters $a$ and $b$. First, we set $b=0.8$ and vary $a$ as $a = 0, 1\times 10^{-3}, 1\times10^{-2}$, and $1\times10^{-1}$. Second , we set $a=1\times 10^{-2}$ and vary $b$ as $ b = 0.1, 0.5, 1$, and $5$. From Fig. \ref{experiment_6_a}, we can see that the optimal $a$ is chosen as the variance of the small coefficients. Smaller $a$ will result in larger misalignment, and larger $a$ will cause slow convergence. From Fig. \ref{experiment_6_b}, we conclude that $b=1$ shows the best performance. Either too large or too small $b$ will result in slower convergence.

\section{Conclusion}

In order to improve the performance of ZA-LMS for near sparse system
identification, an improved algorithm, DWZA-LMS algorithm, is
proposed in this paper by adding a window to the zero-point
attractor in ZA-LMS algorithm and utilizing the magnitude of
estimation error to weight the zero-point attractor. Such
improvement can adjust the zero-point attraction force dynamically
to accelerate the convergence rate with no computational complexity
increased. In addition, the mean square convergence condition,
steady-state MSE and parameter selection of the proposed algorithm
are theoretically analyzed. Finally, computer simulations
demonstrate the improvement of the proposed algorithm and
effectiveness of the analysis.

\appendices
%
%
%


\end{document}